%% file: main.tex
\makeatletter
\def\@copyrightspace{\relax}
\makeatother

\pdfoutput=1

\documentclass[sigconf,authorversion,nonacm]{acmart}

\AtBeginDocument{%
  }

\begin{document}

\title{ConstitutionMaker: Interactively Critiquing Large Language Models by Converting Feedback into Principles}



\author{Savvas Petridis}
\affiliation{%
  \institution{Google Research}
  \city{New York}
  \state{New York}
  \country{USA}
  }
\email{petridis@google.com}

\author{Ben Wedin}
\affiliation{%
  \institution{Google Research}
  \city{Cambridge}
  \state{Massachusetts}
  \country{USA}
  }
\email{wedin@google.com}

\author{James Wexler}
\affiliation{%
  \institution{Google Research}
  \city{Cambridge}
  \state{Massachusetts}
  \country{USA}
  }
\email{jwexler@google.com}

\author{Aaron Donsbach}
\affiliation{%
  \institution{Google Research}
  \city{Seattle}
  \state{Washington}
  \country{USA}
  }
\email{donsbach@google.com}

\author{Mahima Pushkarna}
\affiliation{%
  \institution{Google Research}
  \city{Cambridge}
  \state{Massachusetts}
  \country{USA}
  }
\email{mahimap@google.com}

\author{Nitesh Goyal}
\affiliation{%
  \institution{Google Research}
  \city{New York}
  \state{New York}
  \country{USA}
  }
\email{teshg@google.com}

\author{Carrie J. Cai}
\affiliation{%
  \institution{Google Research}
  \city{Mountain View}
  \state{California}
  \country{USA}
  }
\email{cjcai@google.com}

\author{Michael Terry}
\affiliation{%
  \institution{Google Research}
  \city{Cambridge}
  \state{Massachusetts}
  \country{USA}
  }
\email{michaelterry@google.com}

\renewcommand{\shortauthors}{Petridis et al.}



\begin{abstract}
Large language model (LLM) prompting is a promising new approach for users to create and customize their own chatbots.
However, current methods for steering a chatbot's outputs, such as prompt engineering and fine-tuning, do not support users in converting their natural feedback on the model's outputs to changes in the prompt or model.
In this work, we explore how to enable users to interactively refine model outputs through their feedback, by helping them convert their feedback into a set of principles (i.e. a constitution) that dictate the model's behavior.
From a formative study, we (1) found that users needed support converting their feedback into principles for the chatbot and (2) classified the different principle types desired by users.
Inspired by these findings, we developed ConstitutionMaker, an interactive tool for converting user feedback into principles, to steer LLM-based chatbots.
With ConstitutionMaker, users can provide either positive or negative feedback in natural language, select auto-generated feedback, or rewrite the chatbot’s response; each mode of feedback automatically generates a principle that is inserted into the chatbot’s prompt.
In a user study with 14 participants, we compare ConstitutionMaker to an ablated version, where users write their own principles.
With ConstitutionMaker, participants felt that their principles could better guide the chatbot, that they could more easily convert their feedback into principles, and that they could write principles more efficiently, with less mental demand.
ConstitutionMaker helped users identify ways to improve the chatbot, formulate their intuitive responses to the model into feedback, and convert this feedback into specific and clear principles.
Together, these findings inform future tools that support the interactive critiquing of LLM outputs.
\end{abstract}

\begin{teaserfigure}
\centering
  \includegraphics[width=0.9\linewidth]{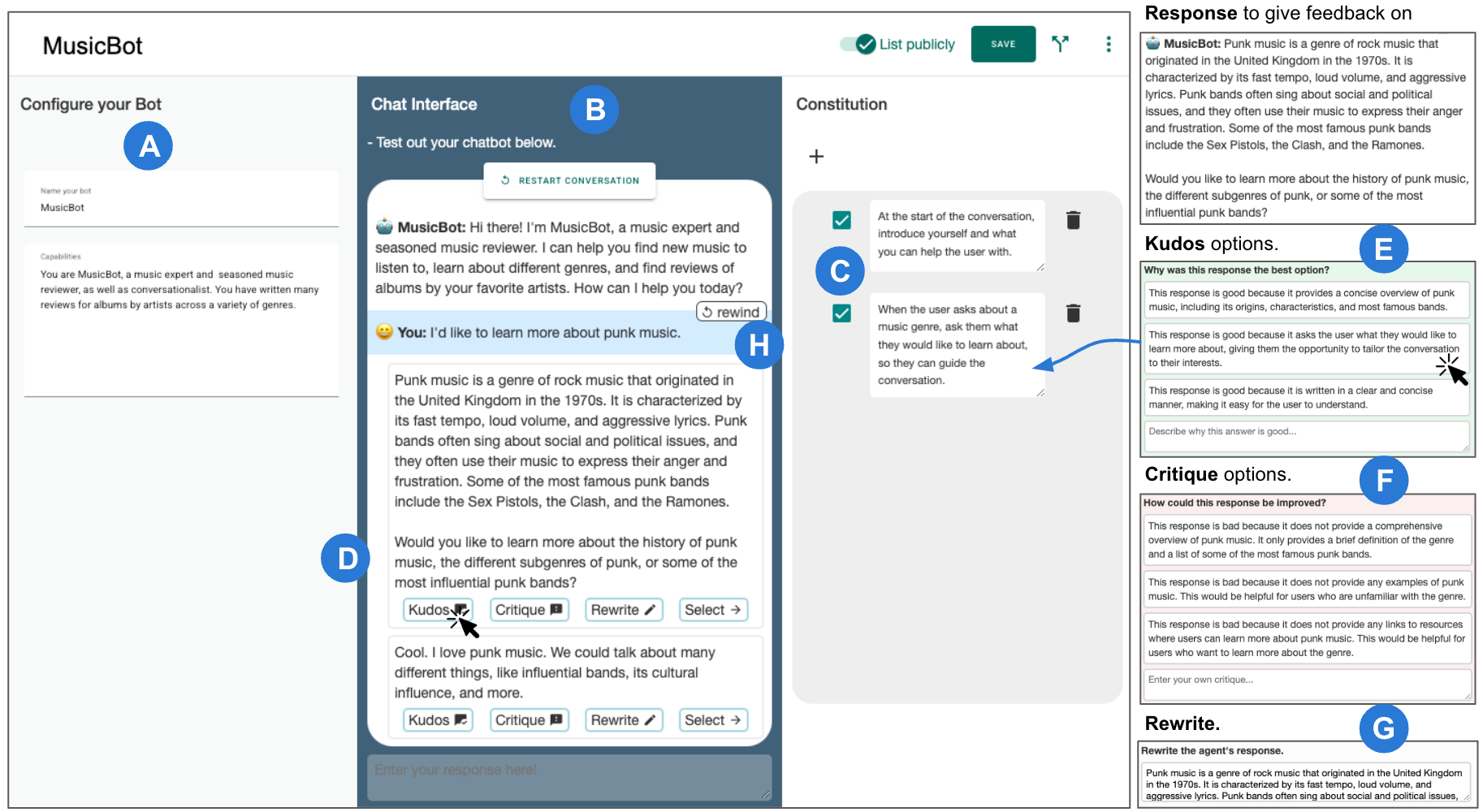}
  \caption{\textit{ConstitutionMaker's Interface.} First, users name and describe the chatbot they'd like to create \textbf{(A)}. 
  ConstitutionMaker constructs a dialogue prompt, and users can then immediately start a conversation with this chatbot \textbf{(B)}.
  At each conversational turn, users are presented three candidate responses by the chatbot, and for each one, three ways to provide feedback: (1) \textit{kudos}, (2) \textit{critique}, and (3) \textit{rewrite}.
  Each feedback method elicits a principle, which gets added to the Constitution in \textbf{(C)}.
  Principles are rules that get appended to the dialogue prompt.
  Giving \textit{kudos} to an output \textbf{(D)} entails providing positive 
  feedback, either through selecting one of three generated positive rationales or by writing custom positive feedback. \textit{Critiquing} \textbf{(F)} is the same but providing negative feedback. And finally, \textit{rewriting} \textbf{(G)} entails revising the response to generate a principle. 
  }~\label{fig:cm-interface}
  \Description{}
\end{teaserfigure}

\maketitle

\input{1_introduction}

\input{2_related_work}

\input{3_formative_study}

\input{4_constitution_maker}

\input{5_implementation}

\input{6_user_study}

\input{7_findings}

\input{8_discussion}

\input{9_conclusion}



\bibliographystyle{ACM-Reference-Format}
\bibliography{main}


\end{document}

%% file: 1_introduction.tex
\section{Introduction}
\begin{figure*}
\centering
  \includegraphics[width=0.8\linewidth]{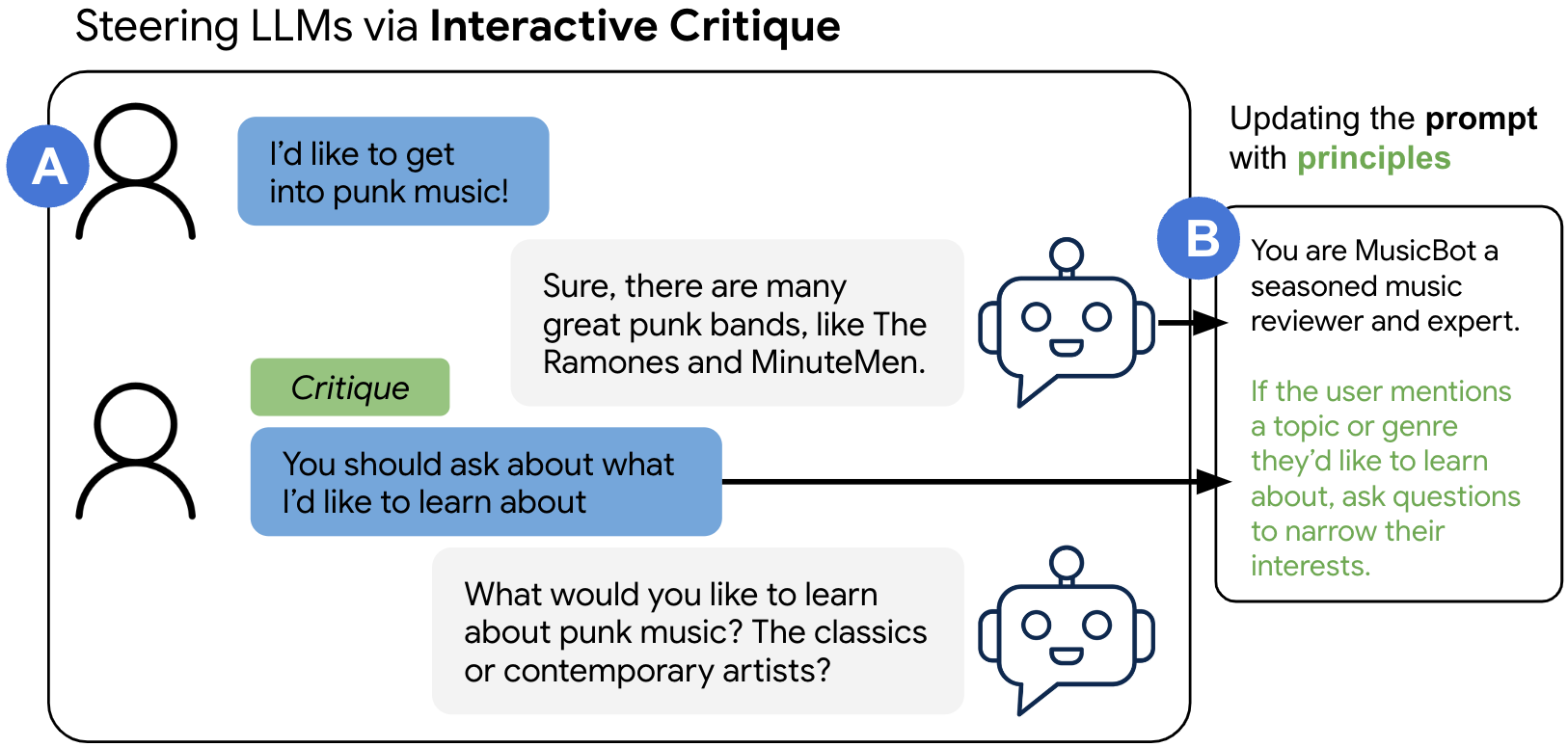}
  \caption{Illustration of steering an LLM via \textbf{interactive critique.} In conversations with LLMs like Chat-GPT and Bard, users provide natural language feedback, as they would to another person, to steer the LLM to better outputs. In this example, the user critiques the following music-recommender LLM to ask questions to establish the user's interests, prior to providing information on a genre \textbf{(A)}. This method of customizing LLMs is currently not persistent; users need to repeat instructions during each new interaction. To help create a persistent form of this customization, this work investigates converting user feedback into \textit{principles} \textbf{(B)}, which are specific rules to direct the LLM's behavior inserted into the LLM prompt.}~\label{fig:interactive-critique}
  \Description{}
\end{figure*}

Large language models (LLMs) can be applied to a wide range of problems, ranging from creative writing assistance \cite{wordcraft,sparks,anglekindling,popblends} to code synthesis \cite{llm-programming-liu,natural_language_programming,genline}.
Users currently customize these models to specific tasks through strategies such as prompt engineering \cite{few-shot-learners}, parameter-efficient tuning \cite{prompt-tuning}, and fine-tuning \cite{fine-tuning}.

In addition to these common methods for customizing LLMs, recent work has shown that users would also like to directly steer these models with \textit{natural language feedback} (Figure \ref{fig:interactive-critique}A).
More specifically, some users want to be able to \textit{critique} the model's outputs to specify how they should be different \cite{sparks_of_agi}.
We call this customization strategy \textbf{interactive critique}.

When interacting with a chatbot like ChatGPT\footnote{https://chat.openai.com/} \cite{gpt} or Bard\footnote{https://bard.google.com}, interactive critique will often alter the chatbot's subsequent responses to conform to the critique.
However, these changes are not persistent: users must repeat these instructions during each new interaction with the model.
Users must also be aware that they can actually alter the model's behavior in this way, and must formulate their critique in a way that is likely to lead to changes in the model's future responses.
Given the potential value of this mode of customizing there is an opportunity to provide first-class support for empowering users to customize LLMs via natural language critique.

In the context of model customization, Constitutional AI \cite{constitutional_ai} offers a specific customization strategy involving natural language \textit{principles}.
A principle can be thought of as a rule that the language model should follow, such as, ``Do not create harmful, sexist, or racist content''.
Given a set of principles, a Constitutional AI system will 1) rewrite model responses that violate principles and 2) fine-tune the model with the rewritten responses. Returning to the notion of interactive critique, one can imagine \textit{deriving} new or refined Constitutional AI principles from users' critiques. These derived principles could then be used to alter an LLM's prompt (Figure \ref{fig:interactive-critique}B) or to generate new training data, as in the original Constitutional AI work.

While this recent work has shown principles can be an explainable and effective strategy to customize an LLM, little is known about the human processes of writing these principles from our feedback.
From a formative study, we discovered that there are many cognitive challenges involved in converting critiques into principles.
To address these challenges, we present ConstitutionMaker, an interactive critique system that transforms users' \textit{model critiques} into \textit{principles} that refine the model's behavior. ConstitutionMaker generates three candidate responses at each conversational turn. In addition to these three candidate responses, ConstitutionMaker provides three \textit{principle-elicitation} features: 1) \textit{kudos}, where users can provide positive feedback for a response, 2) \textit{critique}, where users can provide negative feedback for a response, and 3) \textit{rewrite}, where users can rewrite a given response. From this feedback, ConstitutionMaker infers a \textit{principle}, which is incorporated in the chatbot's prompt.

To evaluate how well ConstitutionMaker helps users write principles, we conducted a within-subjects user study with 14 industry professionals familiar with prompting. Participants used ConstitutionMaker and an ablated version that lacked the multiple candidate responses and the principle-elicitation features. In both cases, their goal was to write principles to customize two chatbots. 
From the study, we found that the two different versions yielded very different workflows. With the ablated version, participants only wrote principles when the bot deviated quite a bit from their expectations, resulting in significantly fewer principles being written, in total.
In contrast, in the ConstitutionMaker condition, participants engaged in a workflow where they scanned the multiple candidate responses and gave kudos to their favorite response, leading to more principles overall.
These different workflows also yielded condition-specific challenges in writing principles.
With the ablated version, users would often under-specify principles; whereas, with ConstitutionMaker, users sometimes overspecified their principles, though this occurred less often.
Finally, both conditions would sometimes lead to an issue where two or more of the principles were in conflict with one another.

Overall, with ConstitutionMaker, participants felt that their principles could better guide the chatbot, that they could more easily convert their feedback into principles, and that they could write principles more efficiently, with less mental demand. 
ConstitutionMaker also supported their thought processes as they wrote principles by helping participants 1) recognize ways responses could be better through the multiple candidate responses, 2) convert their intuition on why they liked or disliked a response into verbal feedback, and 3) phrase this feedback as a specific principle.

Collectively, this work makes the following contributions:
\begin{itemize}
    \item A classification of the kinds of principles participants want to write to steer chatbot behavior.
    \item The design of ConstitutionMaker, an interactive tool for converting user feedback into principles to steer chatbot behavior. ConstitutionMaker introduces three novel principle-elicitation features: \textit{kudos}, \textit{critique}, and \textit{rewrite}, which each generate a principle that is inserted into the chatbot’s prompt.
    \item Findings from a 14-participant user study, where participants felt that ConstitutionMaker enabled them to 1) write principles that better guide the chatbot, 2) convert their feedback into principles more easily, and 3) write principles more efficiently, with less mental demand. 
    \item We describe how ConstitutionMaker supported participants’ thought processes, including helping them identify ways to improve responses, convert their intuition into natural language feedback, and phrase their feedback as specific principles. We also describe how the different workflows enabled by the two systems led to different challenges in writing principles and the limits of principles.
\end{itemize}

Together, these findings inform future tools for interactively refining LLM outputs via interactive critique.

%% file: 2_related_work.tex
\section{Related Work}

\subsection{Designing Chatbot Behavior}
There are a few methods of creating and customizing chatbots.
Earlier chatbots employed rule-based approaches to construct a dialogue flow \cite{rule_based_chatbots, rule-based-chatbots2, rule-based-chatbot-3}, where the user's input would be matched to a pre-canned response written by the chatbot designer.
Later on, supervised machine learning approaches \cite{chatbot_social_media, emotional_chatbot} became popular, where chatbot designers constructed datasets consisting of ideal conversational flows.
Both of these approaches, while fairly effective, require a significant amount of time and labor to implement, from either constructing an expansive rule set that determines the chatbot's behavior or from building a large dataset consisting of ideal conversational flows.

More recently, large language model prompting has shown promise for enabling easier chatbot design.
Large, pre-trained models like Chat-GPT \cite{gpt} can hold sophisticated conversations out of the box, and these models are already being used to create custom chatbots in a number of domains, including medicine \cite{gpt-medical-chatbot}.
There are a few ways of customizing an LLM-based chatbot, including prompt engineering and  fine-tuning.
Prompt engineering involves providing instructions or conversational examples in the prompt to steer the chatbot's behavior \cite{few-shot-learners}.
To more robustly steer the model, users can also fine-tune \cite{prompt-tuning} the LLM with a larger set of conversational examples.
Recent work has shown that users would also like to steer LLMs by \textit{interactively critiquing} its outputs; during the conversation they refine the model's outputs by providing follow-up instructions and feedback \cite{sparks_of_agi}.
In this work, we explore how to support users with this type of model steering: naturally customizing the LLM's behavior through feedback, as they interact with it.

A new approach to steering LLM-based chatbots (and LLMs in general), called Constitutional AI \cite{constitutional_ai} involves writing natural language principles to direct the model.
These principles are essentially rules, such as: ``Do not create harmful, sexist, or racist content''.
Given a set of principles, the Constitutional AI approach involves rewriting LLM responses that violate these principles, and then using these tuples of original and rewritten responses to fine tune the LLM.
Writing principles could be a viable and intuitive way for users to steer LLM-based chatbot behavior, with the added benefit of being able to use these principles later to fine tune the model.
However, relatively little is known about the kinds of principles users want to write, and how we might support users in converting their natural feedback on the model's outputs into principles.
In this work, we evaluate three principle elicitation features that help users convert their feedback into principles to steer chatbot behavior.

\subsection{Helping Users Design LLM Prompts}
While LLM prompting has democratized and dramatically sped up AI prototyping \cite{promptmaker}, it is still a difficult and ambiguous process for users \cite{why_johnny_cant_prompt, herding_ai_cats, beyond_few_shot}; they have challenges with finding the right phrasing for a prompt, choosing good demonstrative examples, experimenting with different parameters, and evaluating how well their prompt is performing. \cite{promptmaker}.
Accordingly, a number of tools have been developed to support prompt writing along these lines.

To help users find a better phrasing for their prompt, automatic approaches have been developed that search the LLM's training data for a more effective phrasing \cite{automatic_prompt_optimization, autoprompt}.
In the text-to-image domain, researchers have employed LLMs to generate better prompt phrasings or keywords for generative image models \cite{opal,3dall-e,promptify,reprompt}.
Next, to support users in sourcing good examples for their prompt, \textit{ScatterShot} \cite{scattershot} suggests underrepresented data to include in the prompt from a dataset, and enables users to iteratively evaluate their prompt with these examples.
Similar systems help users source diverse and representative examples via techniques like clustering \cite{selecting_examples_1} or graph-based search \cite{selecting_examples_2}.
To support easy exploration of prompt parameters, \textit{Cells, Generators, and Lenses} \cite{cells} enables users to flexibly test different inputs with instantiations of models with different parameters.
In addition to improving the performance of a single run prompt, recent work has also investigated the benefits of \textit{chaining} multiple prompts together, to improve performance on more complicated tasks \cite{AI-chains, Promptchainer}.
Finally, tools like \textit{PromptIDE} \cite{prompt_ide}, \textit{PromptAid} \cite{promptaid}, and \textit{LinguisticLens} \cite{linguisticlens} support users in evaluating their prompts, by either visualizing the data it produces, or its performance in comparison to other prompt variations.

This work explores a novel, more natural way of customizing a prompt's behavior through interactive critique. ConstitutionMaker enables users to provide natural language feedback on a prompt's outputs, and this feedback is converted into principles that are then incorporated back into the prompt.
We illustrate the value of helping users update a prompt via their feedback, and we introduce three novel mechanisms for converting users' natural feedback into principles for the prompt.

\subsection{Interactive Model Refinement via Feedback}
Finally, ConstitutionMaker is broadly related to systems that enable users to customize their outputs via limited or underspecified feedback.
For example, programming-by-example tools enable users to provide input-output examples, for which the system generates a function that fits them \cite{programming-by-example-1, programming-by-example-2, programming-by-example-3}.
Input-output examples are inherently ambiguous, potentially mapping to multiple functions, and these systems employ a number of methods to specify and clarify the function with the user.
In a similar process, ConstitutionMaker takes ambiguous natural language feedback on the model's output and generates a more specific principle for the user to inspect and edit.
Next, recommender systems \cite{tasteweights,tastepaths,3d-itemspace} also enable users to provide limited feedback to steer model outputs.
One such system \cite{3d-itemspace} projects movie recommendations on a 2D plane, which users can interactively raise or lower portions of it to affect a list of recommendations; in response to these changes, the system provides representative movies for each raised portion to demonstrate how it has interpreted the user's feedback.
Overall, in contrast to these systems, ConstitutionMaker leverages LLMs to enable users to provide natural language feedback and critique the model in the same way we would provide feedback to another person.

%% file: 3_formative_study.tex
\section{Formative Study}
To understand how to support users with writing principles for chatbots, we conducted (1) a one-hour formative study, where we observed eight industry professionals write principles for chatbots of their choice.
These participants all had prompting experience. Two participants are designers and six are software engineers, all at a large technology company. During the workshop, participants used an early version of ConstitutionMaker, without principle elicitation features. They spent 25 minutes writing principles for their chatbot. Afterwards, we discussed the difficulties they faced while writing principles. Finally, we collected the principles they wrote and classified them to understand the kind of principles they wanted to write.

\subsection{Design Goals}
In this section, we summarize a set of three design goals for ConstitutionMaker we established from the formative workshop and subsequent think-alouds.

\begin{itemize}
    \item[D.1] \textbf{Help users recognize ways to improve the chatbot's responses} by showing alternative chatbot responses. Today’s LLMs are quite sophisticated, and even with just a preamble describing how the bot should behave, the chatbot can hold a convincing conversation. Because of this, participants mentioned that it was sometimes hard to imagine how the chatbot’s responses could be improved. This did not mean, however, that they thought the chatbot’s response was perfect, but instead passable and without any glaring errors. Therefore, to help participants recognize better kinds of responses to steer the chatbot to, our first design goal was to provide multiple candidate responses from the chatbot at each conversational turn. This way, participants can compare them and recognize components they like more than others.
    \item[D.2] \textbf{Help convert user feedback into specific principles} to make principle writing easier. One piece of feedback we got from participants was that writing principles involves a difficult two-step process of first (1) articulating one's feedback on the model's current output, and then (2) converting this feedback into a principle for the LLM to follow. Often, one's initial reaction to the model's output is intuitive, and converting that intuition into a principle for the chatbot to follow can be challenging. In addition, once participants had a particular bit of feedback in mind (e.g., ``I don't like how the chatbot didn't introduce itself''), they were unsure how to phrase their principle. However, in line with prior research \cite{why_johnny_cant_prompt,herding_ai_cats}, they found that more concrete principles that specified what should happen and when (e.g., ``Introduce yourself \textit{at the start of the conversation, and state what you can help with}'') generally led to better results. Thus, our second design goal was to help users go from their initial reaction to the model's output to a specific, clearly written principle to steer the model.
    \item[D.3] \textbf{Enable easier testing of principles} to help users understand how well their principles are steering the chatbot’s behavior. As participants wrote more principles, they wanted ways to test these principles to make sure they worked. The early version of ConstitutionMaker only let users restart the conversation, and did not let users enable or disable principles. Users wanted to test individual principles on certain portions of the conversation, to see if the model was generating the correct content. And so, our last design goal was to enable easier testing of principles.
\end{itemize}

\subsection{Principle Classification}
From the formative workshop and follow up sessions, we collected 79 principles in total and classified them to understand the kinds of principles users wanted to write. These principles correspond to a number of very different chatbots, including a show recommender, chemistry tutor, role playing game manager, travel agent and more. We describe common types of principles below.

\textbf{Principles can be either unconditional or conditional}.
Unconditional principles are those that apply at every conversational turn. 
Examples include: (1) those that define a \textit{consistent personality} of the bot (e.g., ``Act grumpy all the time'' or ``Speak informally and in the first person''), (2) those that place \textit{guardrails} on the conversational content (e.g., ``Don't talk about anything but planning a vacation''), and (3) those that establish a consistent \textit{form} for the bot's responses (e.g., ``Limit responses to 20 words''). 
Meanwhile, a conditional principle only applies when a certain condition is meant.
For example, ``Generate an itinerary after all the information has been collected,'' only applies to the conversation when some set of information has been acquired.
Writing a conditional principle essentially defines a computational interaction; users establish a set of criteria that make the principle applicable to the conversation, and once that set of criteria is met, the principle is executed (e.g., an itinerary is generated).

\textbf{Conditional principles can depend on the entire conversation history, the user's latest response, or the action the bot is about to take.}
For example, ``Generate an itinerary after all the information has been collected'' depends on the entire conversation history to determine if all of the requisite information has been collected.
Similarly the following principle written for a machine learning tutor, ``After verifying a user's goal, provide help to solve their problem,'' depends on the conversation history to identify if the user's goal has been verified.
Meanwhile, the principle ``When the user says they had a particular cuisine the night before, recommend a different cuisine,'' written for a food recommender, pertains just to the latest response by the user.
Finally, the condition can depend on the action the bot is about to take, like ``When providing a list of suggestions, use free text rather than bullet points,'' which applies to any situation when the bot thinks it is appropriate to make suggestions.

\textbf{Conditional principles can be fulfilled in a single or multiple conversational turns.} 
For example, the principle ``At the start of the conversation, introduce yourself and ask a fun question to kick off the conversation'' is fulfilled in a single conversational turn, in which the bot introduces itself.
Similarly, ``Before recommending a restaurant, ask the user for their location'' is also fulfilled in a single turn.
Meanwhile, for a role playing game (RPG) bot that guides the user through an adventure, a participant wrote the following principle: ``When the user tries to do something, put up small obstacles. Don’t let them succeed on the first attempt.''
This principle implies that the bot needs to take action multiple turns prior to being fulfilled (e.g., by first putting up a small obstacle and then subsequently letting the user succeed).
Similarly, for a travel agent bot, a user wrote ``Ask questions one-by-one to get an idea of their preferences,'' which also requires multiple conversational turns prior to fulfillment.

In summary, principles can either be conditional, where they apply when a certain condition is met, or unconditional, where they apply at every conversational step. Conditional principles further break down into those that depend on the entire conversation history, the user's last response, or the action the bot is about to take. And finally, conditional principles are either fulfilled in a single turn or multiple conversational turns.

%% file: 4_constitution_maker.tex
\section{ConstitutionMaker}
Inspired by our findings from the formative studies and workshop, we built ConstitutionMaker, an interactive web tool that supports users in converting their feedback into principles to steer a chatbot's behavior. ConstitutionMaker enables users to define a chatbot, converse with it, and within the conversation, interactively provide feedback to steer the chatbot's behavior.

\subsection{Interface and Walk Through} To illustrate how ConstitutionMaker works, let us consider a music enthusiast, Penelope, who would like to design a chatbot that helps you learn about and find new music, called MusicBot. 
She starts by entering the name of her bot and roughly describing its purpose in the ``Capabilities'' section of the interface (Figure \ref{fig:cm-interface}A).
She then starts a conversation with MusicBot, and after MusicBot's introductory message, she asks to learn about punk music (Figure \ref{fig:cm-interface}B).
Fulfilling our first design goal, \textit{help users recognize ways to improve the bot's responses}, at each conversational turn, ConstitutionMaker provides three candidate responses from the bot (Figure \ref{fig:cm-interface}D) that the user can compare and provide feedback on.
Penelope peruses these candidate responses, and of the three responses, she likes the first one, as it invites the user to continue the conversation with a question at the end.
She now wants to write a principle to help ensure that the chatbot will continue to do this in future conversations.

Fulfilling D.2, \textit{help convert user feedback into principles}, ConstitutionMaker provides three principle-elicitation features to support users in converting their feedback to principles: \textit{kudos}, \textit{critique}, and \textit{rewrite}.
Since Penelope likes the response, she selects \textit{kudos} underneath it (Figure \ref{fig:cm-interface}D), which reveals a menu with three automatically generated rationales on why the response is good, as well as a text field for Penelope to enter her own reason.
After scanning the rationales, she selects the second, as it closely matched her own feedback, and subsequently a principle is automatically generated from that rationale (Figure \ref{fig:cm-interface}C).
The \textit{critique} (Figure \ref{fig:cm-interface}F) and \textit{rewrite} (Figure \ref{fig:cm-interface}G) principle elicitation features work similarly, where Penelope can select a negative rationale or rewrite the model's response to generate a principle respectively.
She then inspects the generated principle, decides that it captured her intention well, and decides not to edit it.

Fulfilling D.3, \textit{enable easy testing of principle}, she can then test to see if the chatbot is following her principle by rewinding the conversation (Figure \ref{fig:cm-interface}H) to get a new set of candidate responses from the model. Ultimately, she decides to continue conversing with MusicBot, exploring different user journeys, and using the principle elicitation features to create a comprehensive set of principles.

%% file: 5_implementation.tex
\section{Implementation}
ConstitutionMaker is a web application and utilizes an LLM \footnote{anonymized for peer review} that is promptable in the same way as GPT-3 \footnote{https://openai.com/api/} or PaLM \footnote{https://developers.generativeai.google/}.
In the following section, we go through the implementation of ConstitutionMaker's key features.

\begin{figure*}
\centering
  \includegraphics[width=1\linewidth]{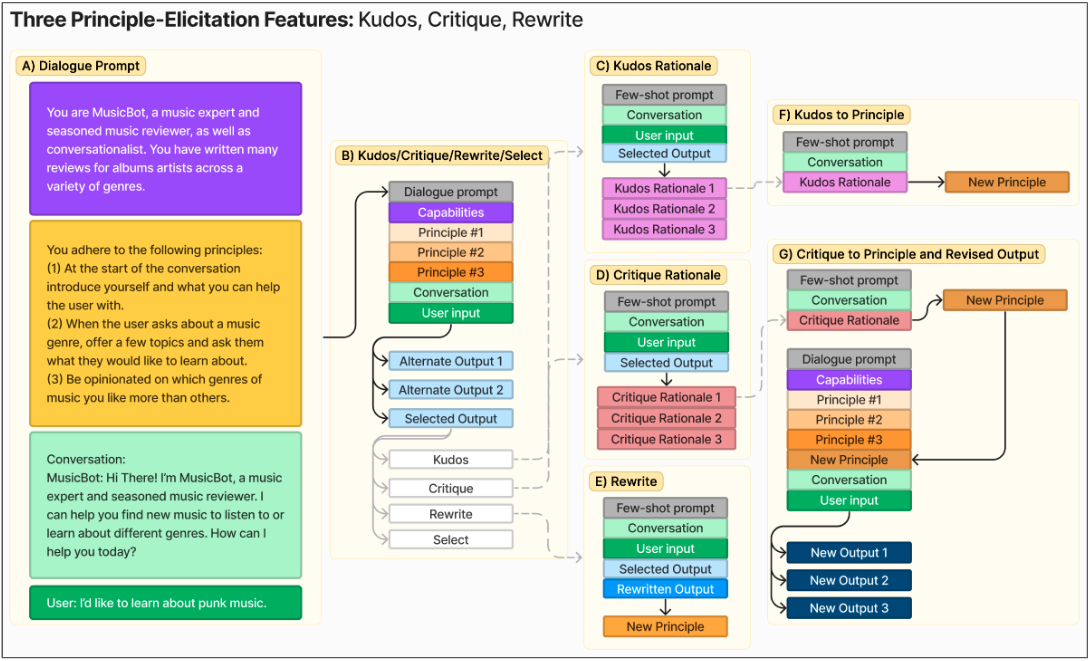}
  \caption{\textbf{The three principle elicitation features: \textit{kudos}, \textit{critique} and \textit{rewrite}}. ConstitutionMaker generates three candidate outputs for the chatbot at each conversational turn, using a dialogue prompt (A) that consists of bot's capabilities (i.e. its purpose), the current set of principles, and the current conversation context. The user can then kudos, critique, or rewrite any of these responses (B). For both kudos and critique, the chosen bot output is then inputted into a few-shot prompt that generates three rationales on why it's good (C) or bad (D), respectively. If the user has given kudos, the rationale is then fed into a subsequent prompt that generates a principle (F) that is then incorporated into the dialogue prompt. Similarly, if the user is critiquing the output, the critique rationale is then fed into a few-shot prompt that generates a new principle, which in turn, is fed back into the dialogue prompt and used to generate a new set of outputs (G). Finally, if the user decides to rewrite, the rewritten output along with the original are sent to a prompt that also generates a principle (E).}~\label{fig:pe-features}
  \Description{}
\end{figure*}

\subsection{Facilitating the Conversation}
To generate the chatbot's response, ConstitutionMaker builds a dialogue prompt (Figure \ref{fig:pe-features}A) behind the scenes.
The dialogue prompt consists of (1) a description of the bot's capabilities, entered by the user (Figure \ref{fig:cm-interface}A), (2) the current set of principles, and (3) the conversation history, ending with the user's latest input.
The prompt then generates the bot's next response, for which we choose the top-3 completions outputted by the LLM to display to users (Figure \ref{fig:pe-features}B).
When the conversation is restarted or rewound, the conversation history within the dialogue prompt is modified; in the case of restarting, the entire history is deleted, whereas for rewinding, everything after the rewind point is deleted.
And finally, if the conversation gets too long for the prompt context window, we remove the oldest conversational turns until it fits.

\subsection{Three Principle Elicitation Features}
All three principle elicitation features output a principle that is then incorporated back into the dialogue prompt (Figure \ref{fig:pe-features}A) to influence future conversational turns.
Giving kudos and critiquing a bot's response consist of a similar process.
For both, the selected bot output is fed into a few-shot prompt that generates rationales, either positive (Figure \ref{fig:pe-features}C) or negative (Figure \ref{fig:pe-features}D).
The user's selected rationale (or their own written rationale) is then sent to a few-shot prompt that converts this rationale into a principle (Figure \ref{fig:pe-features}F and \ref{fig:pe-features}G).
This few-shot prompt leverages the conversation history to create a specific, conditional principle.
For example, for MusicBot, if the critique is ``The bot did not ask questions about the user's preferences,'' a specific, conditional principle might be ``Prior to giving a music recommendation, ask the user what genres or artists they currently listen to.''
Next, for critiques, after the principle is inserted into the dialogue prompt, new outputs are generated to show to the user (Figure \ref{fig:pe-features}G).
Finally, for rewriting the bot's response, we leverage a chain-of-thought \cite{chain_of_thought} style prompt that first generates a ``thought,'' which reasons how the original and rewritten outputs are different from each other, and then generates a specific principle based on that reasoning.
Constructing the prompt with a ``thought'' portion led to principles that captured the difference between the two outputs better than our earlier versions without it.

%% file: 6_user_study.tex
\section{User Study}
To understand (1) if the principle elicitation features help users write principles to steer LLM outputs and (2) what other kinds of feedback they wanted to give, we conducted a 14-participant within-subjects user study, comparing ConstitutionMaker to an ablated version without the principle elicitation features. This ablated version still offered users the ability to rewind parts of the conversation, but participants could only see one chatbot output at a time and had to write their own principles.

\begin{table}
 \begin{tabular}{p{0.2\linewidth} | p{0.75\linewidth}}
\toprule
\textbf{Measure}          & \textbf{Statement (7-point Likert scale)}                                                                                                                      \\ \midrule
\textbf{Effectively Guide}      & Q1. With \{Tool A/B\}, I feel like I was able to write rules that can effectively guide the bot to produce my desired outcomes.                                                  \\[0.1cm]
\textbf{Diversity}        & Q2. With \{Tool A/B\}, I feel like I can think of more diverse rules that can guide the bot in a number of different ways and situations.                                                                   \\[0.1cm]
\textbf{Ease}     & Q3. With \{Tool A/B\}, I felt like it was easy to convert my thoughts and feedback on the bot’s behavior into rules for the bot to follow. \\[0.1cm]
\textbf{Efficiency} & Q4. With \{Tool A/B\}, I felt like I could quickly and efficiently write rules for the bot.                                                                                             \\[0.1cm]
\textbf{Mental Demand}        & Q5. With \{Tool A/B\}, I had to work very hard (mentally) to think of and write rules.                                                           \\ \bottomrule
\end{tabular}
\caption{\textbf{Post-task questionnaire} filled out by participants after they wrote principles for two chatbots, one with both PromptInfuser and the other, with the ablated version. Each statement was rated on a 7-point Likert scale.}
\Description{}
\label{tab:questionnaire}
\end{table}

\subsection{Procedure}
The overall outline of this study is as follows: (1) Participants spent 40 minutes writing principles for two separate chatbots, one with ConstitutionMaker (20 minutes) and the other with the baseline version (20 minutes), while thinking aloud. Condition order was counterbalanced, with chatbot assignment per condition also balanced. (2) After writing principles for both chatbots, participants completed a post-study questionnaire, which compared the process of writing principles with each tool. (4) Finally, in a semi-structured interview, participants described the positives and negatives of each tool and their workflow. The total time commitment of the study was 1 hour.

The two chatbots participants wrote principles for were VacationBot, an assistant that helps users plan and explore different vacation options, and FoodBot, an assistant that helps you plan your meals and figure out what to eat.
These two chatbots were chosen because they support tasks that most people are experienced with, so that participants could have opinions on their outputs and write principles.
For both chatbots, participants were given the name and capabilities (Figure \ref{fig:cm-interface}A), so as to focus predominantly on principle writing.
Also, we balanced which chatbot went with each condition, so half of the participants used ConstitutionMaker to write principles for VacationBot, and the other half used the baseline for VacationBot.
Finally, prior to using each version, participants watched a short video showing that respective tool's features.

To situate the task, participants were asked to pretend to be a chatbot designer and that they were writing principles to dictate the chatbot's behavior so that it performs better for users.
We wanted to observe their process for writing principles and see if the tools impacted how many principles they could write, so we encouraged participants to write at least 10 principles for each chatbot, to give them a concrete goal and to motivate them to write more principles.
However, we emphasized that this was only to encourage them to write principles and that they should only write a principle if they thought it would be useful to future users.

\subsection{Measurements and Analysis}

\subsubsection{Questionnaire} We wanted to understand if and how well ConstitutionMaker's principle elicitation features help users write principles. Our questionnaire (Table \ref{tab:questionnaire}) probes a few aspects of principle writing including participants' perception of (1) how well the output principles \textit{effectively guide} the bot, the \textit{diversity} of the output principles, how \textit{easy} it was to convert their feedback into principles, the \textit{efficiency} of their principle writing process, and the requisite \textit{mental demand} \cite{nasa-tlx} for writing principles with each tool.
To compare the two conditions, we conducted paired sample Wilcoxon tests with full Bonferroni correction, since the study was within-subjects and the questionnaire data was ordinal.

\subsubsection{Feature Usage Metrics}
To shed further light on which tool helped participants write principles more, we recorded the number of principles written for each condition. Moreover, to understand which of the principles elicitation features was most helpful, we recorded how often each was used during the experimental (full ConstitutionMaker) condition. To compare the average number of principles collected across the two conditions, we conducted a paired t-test.

\subsection{Participants}
We recruited 14 industry professionals at a large technology company (average age = 32, 6 female and 8 male) via an email call for participation and word of mouth.
These industry professionals included UX designers, software engineers, data scientists, and UX researchers.
Eligible participants were those that had at least written a few LLM prompts in the past.
The interviews were conducted remotely.
Participants received a \$25 gift card for their time.

%% file: 7_findings.tex
\section{Findings}
\begin{figure*}
\centering
  \includegraphics[width=0.8\linewidth]{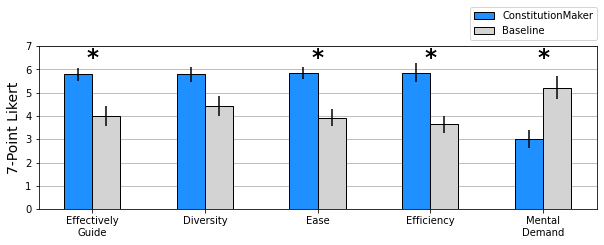}
  \caption{\textbf{Questionnaire results comparing the two conditions.} Bars are standard error and an asterisk indicates a statistically significant difference (after full Bonferroni correction). See Table \ref{tab:questionnaire} for the corresponding question for each of these measures. }~\label{fig:quant_results}
  \Description{This is bar chart comparing ConstitutionMaker to the baseline condition in the user study, across all the measures from the post-task questionnaire. The y-values are from 1 to 7 for a Likert scale. In all categories, ConstitutionMaker scored higher than the baseline, but this difference was significant for: Effectively guide, Ease, Efficiency, and Mental Demand. These were indicated by asterisks.}
\end{figure*}

\textit{Quantitative Findings}. From the exit interviews, 12 of 14 participants \textbf{preferred} ConstitutionMaker to the baseline version.
The results from the questionnaire are summarized in Figure \ref{fig:quant_results}.
We found that ConstitutionMaker was perceived to be more helpful for writing rules that \textbf{effectively guided} the bot (Z = 8, p = .007), scoring on average 5.79 ($\sigma$ = 1.01) whereas the baseline scored 4.0 (1.51). When participants rewound parts of the conversation, they thought the ConstitutionMaker principles were followed by the bot better.
Next, participants felt it was significantly \textbf{easier} (Z = 10.5, p = .006) to convert their feedback into principles with ConstitutionMaker ($\mu$ = 5.86, $\sigma$ = 0.91) than with the baseline ($\mu$ = 3.93, $\sigma$ = 1.39).
The automatic conversion of kudos and critiques into principles eased the process of converting intuitive feedback into clear criteria for the bot to follow.
Participants perceived that they were significantly more \textbf{efficient} (Z = 5, p = .004) writing rules with ConstitutionMaker ($\mu$ = 5.86, $\sigma$ = 1.51) than with the baseline ($\mu$ = 3.64, $\sigma$ = 1.34).
Finally, participants also felt that writing principles with ConstitutionMaker ($\mu$ = 3.0, $\sigma$ = 1.41) required significantly less \textbf{mental demand} (Z = 1.5, p = .002) than with the baseline ($\mu$ = 5.21, $\sigma$ = 1.78).
There was no statistically significant difference for \textbf{diversity} (Z = 19, p = .06); participants felt that they exercised their creativity and wrote relatively diverse principles in the baseline.

Next, regarding the feature usage metrics, participants wrote significantly \textbf{more principles} (t(13)=4.73, p < .001) with ConstitutionMaker than with the baseline; participants wrote on average 6.78 ($\sigma$ = 2.11) principles per chatbot with ConstitutionMaker and 4.42 ($\sigma$ = 1.24) principles with the baseline.
Of the 95 principles written in the ConstitutionMaker condition, 40 (42.1\%) came from kudos, where 37 were selected from the generate rationales and 3 were written; 28 (29.5\%) came from critique, where 8 were selected and 20 were written; 13 (13.6\%) came from the rewrite features; 14 (14.7\%) were manually written.
Participants found rewriting a bit cumbersome and preferred the less intensive workflow of describing what they liked or did not like to generate principles.
In the following sections, we provide further context to these findings.

\subsection{Participants' Workflows for Writing Principles}
The two conditions led to quite different workflows for writing principles.
In the ConstitutionMaker condition, participants commonly scanned the three candidate outputs from the chatbot, identified the output they liked the most, and then gave kudos to that output if they thought it had a quality that was not currently reflected in their principles.
For example, while P1 was working on FoodBot, he was asking for easy-to-make vegetarian dishes, and in one of the bot's candidate outputs, each suggested dish had a short description and explanation on why it was easy.
He appreciated this, skimmed the kudos, and then selected one that conveyed this positive feature of the bot's output.
However, if participants disliked all of the responses, they would then switch to critiquing one of the outputs.
Accordingly, this kudos-first workflow helps to explain how most of the principles participants wrote came from that principle elicitation feature.

Meanwhile, in the baseline condition, participants generally wrote principles when the bot deviated quite a bit (in a negative way) from what they expected.
P8 explained, ``\textit{Here it feels like what I more naturally do is write corrective rules, to guardrail anything that goes weird...If it’s already doing the right thing it doesn't need a rule from me. I wouldn't feel the need to write those.}''
In the baseline condition, participants only see one candidate output from the chatbot, and this might deemphasize the stochastic nature of the chatbot's outputs.
As a result, when participants are okay with a response, they could feel that they do not need to write a principle to encourage that kind of response further.
Overall, with the baseline, participants predominantly wrote principles to steer the LLM from less optimal behavior, while with ConstitutionMaker, participants mostly used kudos the most to encourage behavior they liked.

\subsection{ConstitutionMaker Supported Users' Thought Processes}
In the following section, we discuss how ConstitutionMaker supported participants' thought process from (1) forming an intuition on ways the chatbot could be improved, (2) expressing this intuition as feedback, and (3) converting this feedback into a specific and clear principle.

\subsubsection{Multiple chatbot outputs helped participants form an intuition on how the model could be steered.}
As P5 was using the baseline after using ConstitutionMaker, she explained how she wished she could see multiple outputs again: ``\textit{Sometimes, I don't know what I'm missing [in the baseline]. I'm thinking of the Denali hiking example [which occurred when she wrote principles for VacationBot with ConstitutionMaker]. Two of the responses didn't mention that Denali was good for young children. But one did, and I was able to pull that out as a positive principle.}''
While she was writing principles for VacationBot with ConstitutionMaker, P5 started off the conversation saying she was looking for suggestions for her family, which included two young children.
As the conversation progressed and a general location was established, P5 then asked for hiking recommendations, for which the bot gave some, but only one of its responses highlighted that the hikes it was recommending were good for young children.
P5 gave kudos to that response and created the following principle: ``Consider information previously inputted by the user when providing recommendations.''
Ultimately, it can be hard to form opinions on responses without getting exposed to alternatives, so by providing multiple chatbot outputs, ConstitutionMaker supported participants in forming an intuition on how the model might be steered.

\subsubsection{Automatically providing kudos and critique rationales helped participants formulate their intuitive feedback.}
Upon seeing a candidate response from the chatbot, participants could intuitively tell if they liked or disliked it, but struggled to articulate their thoughts.
The automatically generated kudos and critiques helped participants recognize and formulate this feedback.
For example, while working on FoodBot with ConstitutionMaker, P9 asked the bot to identify the pizzeria with the best thin crust from a list of restaurants provided in a prior turn.
The bot responded with, ``Pizzaiolo has the best thin crust pies.''
P9 knew he did not like the response, so he went to critique it and selected the following generated option: ``This response is bad because it does not provide any information about the other pizza places that the user asked about.''
The following principle was generated: ``If the user asks about a specific attribute of a list of items, provide information about all of the items in the list that have that attribute,'' which then produced a set of revised responses that compared the qualities of each pizzeria's crusts.
Reflecting on this process, P9 stated, ``\textit{I didn't like that last answer [from FoodBot], but I didn't have a concrete reason why yet...I didn't really know how to put it into words yet...but reading the suggestions gave me at least one of many potential reasons on why I didn't like the response.}''
Thus, ConstitutionMaker helped participants transition from \textit{fast thinking} \cite{fast_slow_thinking}, that is, their intuitive and unconscious responses to the bot's outputs, to \textit{slow thinking}, a more deliberate, conscious formulation of their feedback.

\subsubsection{Generating principles from feedback helped users write clear, specific principles}
Sometimes the generated kudos and critique rationales did not capture participants' particular feedback on the chatbot, and so they would then write their own.
Their feedback was often under-specified, and ConstitutionMaker helped convert this feedback into a clear and specific principle.
For example, P4 was writing principles for VacationBot using ConstitutionMaker.
During the conversation, they had told VacationBot that they were planning a week long vacation to Japan, to which the bot immediately responded with a comprehensive 7-day itinerary.
P4 then wrote in their own critique: ``This response is bad because it does not take into account the user's interests.''
The resulting principle was, ``When the user mentions a location, ask them questions about what they are interested in before providing an itinerary.''
This principle was aligned with what P4 had in mind, and reflecting on her experience using ConstitutionMaker, she stated, ``\textit{When I would critique or kudos, it would give examples of principles that were putting it into words a little bit better than I could about like what exactly I was trying to narrow down to here.}''
Finally, even when the resulting principle was not exactly what they had in mind, participants appreciated the useful starting point it provided.
Along these lines, P11 explained, ``\textit{It was easier to say yes-and with Tool A [ConstitutionMaker]. Where it [the generated principle] wasn't all the way there, but I think it's 50\% of the way there, and I can get it to where I want to go.}''
Overall, ConstitutionMaker helped participants specify their feedback into clear principles.

\subsection{Participants' Workflows Introduced Challenges with Writing Principles}
Participants struggled to find the right level of granularity for their principles, and the two conditions led to different problems in this regard.
Both workflows had participants switch roles from \textit{end-user}, where participants experimented with different user journeys, to \textit{bot designer}, where they evaluated the bot's responses to write principles.
The more conversation-forward interface of the baseline blurred the distinction between these two roles.
P3 explained that without the multiple bot outputs and principle elicitation features, ``\textit{you can simulate yourself as the user a lot better in this mode [the baseline]}.''
And by leaning further into this user role, participants wrote principles that were more conversational, but under-specified.
For example, while writing principles for FoodBot with the baseline, P11 wrote the principle ``Be cognizant of the user's dietary preferences.''
What P11 really had in mind was a principle that specified that the bot should ask the user for their preferences and allergies prior to generating a meal plan.
These underspecified principles often did not impact the bot's responses and would frustrate participants while they used the baseline.

Meanwhile, while using ConstitutionMaker, the opposite problem occurred, where users' workflows led to principles that were \textit{over}-specified.
For example, while working on VacationBot, P7 asked the model to help him narrow down a few vacation options, and the model proceeded to ask them questions (without any principle written specifying so).
Appreciating that the model was gathering context, they selected a kudos that praised the model for asking about the user's budget constraints prior to recommending a vacation destination.
The resulting principle was, ``Ask the user their budget before providing vacation options.''
However, once this principle came into effect, the model's behavior then anchored specifically to only asking for budget prior to making a recommendation.
And so, this workflow of providing feedback at every conversational step, instead of for entire conversations, led to a principle that was too specific and impacted the bot's performance negatively.
While users generally appreciated ConstitutionMaker's ability to form specific principles from their feedback, there were rare instances where the principles were too specific.

Finally, in both conditions, by switching back and forth between the end-user to bot designer roles, participants would sometimes write principles that conflicted with each other.
For example, while P2 was working on VacationBot with the baseline, they asked the bot for dog-friendly hotel recommendations in the Bay Area, and VacationBot responded with three recommendations.
P2 wanted more recommendations and wrote a principle to ``Provide >= 10 recommendations.''
Later on in the conversation, P2 now had a list of dog-friendly restaurants, with their requisite costs, and he asked VacationBot which it recommends, to which it responded by listing positive attributes of all the hotel options.
P2, who now wanted a decisive, single response wrote the following principle: ``If I ask for a recommendation, give *1* recommendation only.''
VacationBot, now with two conflicting principles on the number of recommendations to provide, alternated between the two.
Ultimately, by providing feedback on individual conversational turns, participants ended up with conflicting principles.
P8 imagined a different workflow, where he would experiment with full user journeys and then write principles: ``\textit{I think it might help me to actually go through the [whole] user flow and then analyze it as a piece instead of switching...it would allow me to inhabit one mindset [either bot designer or user] for a period of time and then switch mindsets.}''
In summary, one's workflow as they probe and test the model impacts the types of principles they produce and the challenges they face.

\subsection{The Limits of Principles}
Some participants questioned if writing natural language principles was the optimal way to steer all aspects of a bot's behavior. 
While writing a principle to shorten the length of the chatbot's responses, P13 reflected, ``\textit{It feels a little weird to use natural language to generate the principles...it doesn't feel efficient, and I'm not sure how it's going to interpret it.}''
They imagined that aspects like the \textit{form} of the model's responses would be better customized with UI elements like sliders to adjust the length of the bot's responses, or exemplifying the structure of the bot's response (e.g., an indented, numbered list for recommendations) for the model to follow, instead of describing these requests in natural language.
In a similar vein, P14 noticed that her principles pertained to different parts of the conversation, and as a list, they seemed hard to relate to each other.
She wanted to structure and provide feedback on higher-level ``conversational arcs,'' visually illustrating the flow and ``forks in the road'' of the conversation (e.g., ``If the user does X, do Y. Otherwise, do Z'').
Principles are computational in a sense, and they dictate the ways the conversation can unfold; there might be better ways to let users author this flow, other than with individual principles.

%% file: 8_discussion.tex
\section{Discussion}

\subsection{Supporting Users in Clarifying and Iterating Over Principles}
Finding the right granularity for a principle was sometimes challenging for participants; they created under-specified principles that did not impact the conversation, as well as over-specified principles that applied too strictly to certain portions of the conversation.
One way to support users in finding the right granularity could be generating questions to help them reflect on their principle.
For example, if an abstract principle is written (e.g., ``Ask follow up questions to understand the users preferences''), an LLM prompt can be used to pose clarifying questions, such as, ``What kind of follow up questions should be asked?'' or, ``Should I do anything differently, depending on the user's answers?''
Users could then answer these questions the best they can and then the principle could be updated automatically.
Alternatively, another way to help users reflect on their principles might be to engage in a side conversation with a chatbot to help clarify them.
This chatbot could pose similar questions as those suggested above, but it might also provide examples of chat snippets that adhere to or violate the principle.
As they converse, the chatbot might continue to pose clarifying questions, while the principle is updated on the fly.
Thus, future work could examine supporting users in interactively reflecting upon and clarifying their principles.

\subsection{Organizing Principles and Supporting Multiple Principle Writers}
As participants accumulated principles, it was increasingly likely that there was a conflict between two of them, and it was harder for them to get an overview of how their principles were affecting the conversation.
One way to prevent potential principle conflicts is to leverage an LLM to conduct a pairwise comparison of principles to assess if any two are at odds, and then suggest a solution.
This kind of conflict resolution, while useful for a single principle writer, would be crucial in cases when multiple individuals are writing principles to improve a model.
Multiple principle writers would be useful for gathering feedback to improve the overall performance of the model, but with many generated principles, it is increasingly important to understand how they might impact the conversation.
Perhaps a better way to prevent conflicts is to organize principles in a way that summarizes their impact on the conversation.
Principles are small bits of computation; there are conditions when they are applicable, and depending on those conditions, the bot's behavior might branch in separate directions.
One way to organize principles is to construct a state diagram, which illustrates the potential set of supported user flows and bot responses.
With this overview, users could be made better aware of the overall impact of their principles, and they could then easily revise it to prevent conflicts.
Therefore, another rich vein of future work is developing mechanisms to resolve conflicts in larger sets of principles, as well as organizing them into an easily digestible overview.

\subsection{Automatically Ideating Multiple User Journeys to Support Principle Writing}
To test the chatbot and identify instances where the model could be improved, participants employed a strategy where they went through different user journeys with the chatbot.
Often their choice of user journeys were biased toward what they were interested in, or they only tested the most common journeys.
One way to enable more robust testing of chatbots could be to generate potential user personas and journeys to inspire principle writers.
Going further, these generated user personas could then be used simulate conversations with the chatbot being tested \cite{park2023generative}.
For example, for VacationBot, one user persona might be a parent looking for nearby, family friendly vacations.
A dialogue prompt could be generated for this persona and then VacationBot and this test persona could converse for a predefined set of conversational turns.
Afterwards, users could inspect the conversation, and edit or critique VacationBot's responses to generate principles.
This kind of workflow could sidestep the challenge of repeatedly shifting from an end-user's perspective to a bot-designer's perspective, which exists in the current workflow.
At the same time, users would be able to evaluate fuller conversational arcs, as opposed to single conversational turns.
Thus, another line of future work is supporting users in exploring diverse user journeys with their chatbot, as well as exploring workflows that require less perspective switching.

\subsection{Limitations and Future Work}
A set of well-written principles is often not enough to robustly steer an LLM's outputs. As more principles are written, an LLM might ``forget'' to apply older principles \cite{herding_ai_cats}. This work focuses on helping participants convert their intuitive feedback into clear principles, and we illustrate that the principle-elicitation features help with that process. However, in line with the original Constitutional AI workflow \cite{constitutional_ai}, future work can focus on using these principles to generate a fine-tuning dataset, so that the model robustly follows them.

Next, while we selected chatbots for two very common use cases for the study (vacation and food), participants might not have been very knowledgeable or opinionated in these areas. Future work can explore how these principle-elicitation features help users when writing principles for chatbot use cases that they are experts in. That being said, it was necessary to choose two chatbot use cases for the study to enable a fair comparison across the two conditions.

%% file: 9_conclusion.tex
\section{Conclusion}
This paper presents ConstitutionMaker, a tool for interactively refining LLM outputs by converting users' intuitive feedback into principles.
ConstitutionMaker's design is informed by a formative study, where we also collected and classified the types of principles users wanted to write.
ConstitutionMaker incorporates three principle-elicitation features: kudos, critique, and rewrite.
In a user study with 14 industry professionals, participants felt that ConstitutionMaker helped them (1) write principles that \textit{effectively guided} the chatbot, (2) convert their feedback into principles more \textit{easily}, and (3) write principles more \textit{efficiently}, with (4) less \textit{mental demand} than the baseline.
This was due to ConstitutionMaker supporting their thought processes, including helping them to: identify ways to improve the bot's responses, convert their intuition into verbal feedback, and phrase their feedback as specific principles.
There are many avenues of future work, including supporting users in iterating on and clarifying their principles, organizing larger sets of principles and supporting multiple writers, and helping users test chatbots across multiple user journeys.
Together, these findings inform future tools that support interactively customizing LLM outputs.